\documentclass[12pt]{article}

\usepackage{epsfig,amsmath,amssymb,latexsym}

\providecommand{\beqa}{\begin{eqnarray}}
\providecommand{\eeqa}{\end{eqnarray}}

\def\Orb{{\mathbf{S}^1/\mathbf{Z}_2}}
\def\Z2{{\mathbf{Z}_2}}
\def\mX{{\mathbf{X}}}

\numberwithin{equation}{section}

\begin{document}

\thispagestyle{empty}
\rightline{LMU-ASC 03/07}

\begin{center}

{\bf\Large Supersymmetry Breaking with Zero Vacuum Energy}\\
\vspace{.3truecm}
{\bf\Large in M-Theory Flux Compactifications}

\vspace{1.3truecm}

Axel Krause\footnote{Axel.Krause@physik.uni-muenchen.de}

\vspace{.7truecm}

{\em Arnold Sommerfeld Center for Theoretical Physics\\
Department f\"ur Physik, Ludwig-Maximilians-Universit\"at M\"unchen\\
Theresienstr.~37, 80333 M\"unchen, Germany}

\end{center}

\vspace{1.0truecm}


\begin{abstract}

An attractive mechanism to break supersymmetry in vacua with zero
vacuum energy arose in $E_8\times E_8$ heterotic models with hidden sector gaugino condensate. An $H$-flux balances the exponentially small condensate on shell and fixes the complex structure moduli. At quantum level this balancing is, however, obstructed by the quantization of the $H$-flux. We show that the warped flux compactification background in heterotic M-theory can solve this problem through a warp-factor suppression of the integer flux relative to the condensate. We discuss the suppression mechanism both in the M-theory and the 4-dimensional effective theory and provide a derivation of the condensate's superpotential which is free of delta-function squared ambiguities.

\end{abstract}


\newpage
\pagenumbering{arabic}

\section{Introduction}

An old outstanding problem in superstring theory is the breaking of supersymmetry with zero or very little generation of vacuum energy. A mechanism achieving this had been proposed early on in
\cite{Dine:1985rz} in the context of Calabi-Yau compactifications of the heterotic superstring. The idea was to utilize the perfect square potential \cite{Dine:1985rz}, \cite{Bergshoeff:1989de}
\beqa
S = -\frac{1}{2\kappa_{10}^2}
\int_{\phantom{}_{\mathbf{R^{1,3}}\times\mathbf{X}}}
\hspace{-8mm} e^{-\phi}
\Big(H-\frac{\alpha'}{16}e^{\phi/2}
\text{tr} \bar\chi \Gamma^{(3)} \chi \Big)^2
\label{HetStringSquare}
\eeqa
which is built from the Neveu-Schwarz three-form flux $H$ and the gaugino condensate $\langle \text{tr} \bar\chi \Gamma^{(3)} \chi\rangle$ in the hidden gauge sector \cite{Ferrara:1982qs}. Here, $\omega^2 = \omega\wedge\star\omega$, $\phi$ is the dilaton and $\mathbf{X}$ a Calabi-Yau threefold. The condensate can only assume values proportional to the Calabi-Yau's holomorphic three-form $\Omega$ and its complex conjugate. Due to its positivity, this potential minimizes at zero vacuum energy by equating the gaugino condensate with a non-zero $H$-flux of Hodge type $H^{(0,3)}$ and $H^{(3,0)}$ which breaks supersymmetry. Shortly thereafter, it was discovered, however, that the flux appears always quantized in string units \cite{Rohm:1985jv}
\beqa
\frac{1}{2\pi\alpha'} \int_{\Sigma_3} H
= 2 \pi N \; , \quad N \in \mathbf{Z}
\label{Quant}
\eeqa
with $\Sigma_3 \in H_3(\mX,\mathbf{Z})$ being an arbitrary 3-cycle.
This quantum feature, which ensures single-valuedness of the partition function in the presence of worldsheet instantons, clearly spoils the continuous matching between $H$-flux and gaugino condensate. With the condensate scale chosen to be of order $\Lambda = 10^{13}\,\text{GeV}$ to endow the gravitino with a TeV scale mass and the $H$-flux adopting values at the string-scale, no matching is possible and the best one can do is to set the flux to zero, remaining with a large uncanceled vacuum energy from the condensate.

On a Calabi-Yau with non-trivial fundamental group one can have in addition Wilson lines with gauge Chern-Simons contributions to $H$. The gauge invariant field-strength, $H = dB + \frac{\alpha'}{4} (\omega_L-\omega_{YM})$, includes gauge and Lorentz Chern-Simons terms as well, whereas the quantization condition (\ref{Quant}) applies only to $H=dB$. Hence setting $dB$ and thus $N$ to zero, one might try to compensate the gaugino condensate in the perfect square with a Chern-Simons term \cite{Derendinger:1986}. This has been studied recently in \cite{Gukov:2003cy} (see also \cite{Curio:2005ew}). One has to notice, however, that in this case \beqa
\frac{1}{2\pi\alpha'} \int_{\Sigma_3} H = \frac{1}{2\pi\alpha'} \int_{\Sigma_3} \frac{\alpha'}{4}(\omega_L-\omega_{YM})
\eeqa
is only well defined modulo integers \cite{Rohm:1985jv}.

We are proposing an essentially model-independent dynamical solution to this problem which drives the theory dynamically to the heterotic M-theory rather than the weakly coupled heterotic superstring regime, fixing the dilaton close to the phenomenologically preferred critical orbifold length. The warped heterotic M-theory flux compactification background plays a key role. Its warp-factor enters the flux and gaugino condensate parts of the perfect square unequally, with the net result of suppressing the quantized flux against the condensate. The warp-factor being a {\em continuous} parameter, which depends on the dynamical orbifold length, will allow for an exact dynamical minimization of the perfect square even though the flux is quantized in discrete units. Hence, we obtain a dynamical mechanism giving zero vacuum energy with fixed dilaton and broken supersymmetry.

\section{Warped M-Theory Flux Geometries}

Since heterotic M-theory flux compactifications will play such an important role for the successful flux-condensate
balancing, let us first collect a few relevant facts about them.
The two boundaries of heterotic M-theory are separated along the eleventh direction and are located at $x^{11}=0$ (visible) and $x^{11}=L$ (hidden). They represent magnetic sources for the four-form $G$-flux causing the background geometry to be warped.
The general warped background for an $N=1$ compactification
of this theory on $\mX\times\Orb$ has been derived in \cite{Curio:2000dw}, \cite{Krause:2001qf}, \cite{Curio:2003ur} (the linearized approximation in \cite{Witten:1996mz}). Its metric reads
\beqa
ds^2 = \hat g_{MN} dx^M dx^N = e^{2b(x^{11})} g_{\mu\nu} dx^\mu dx^\nu
+ e^{2f(x^{11})} g(\mathbf{X})_{lm} dy^l dy^m + e^{2k(x^{11})}
dx^{11} dx^{11}
\label{Back}
\eeqa
and the three warp-factors depend only on the $\Orb$ orbifold coordinate. Supersymmetry requires that
\beqa
b(x^{11}) = -f(x^{11})
\label{Rel1}
\eeqa
and we can use the coordinate reparameterization freedom of $x^{11}$ to set
\beqa
k(x^{11}) = f(x^{11}) \; .
\label{Rel2}
\eeqa
without loss of generality. The solution to $f(x^{11})$ is then given by
\beqa
e^{3f(x^{11})} = |1-x^{11}Q_v|
\eeqa
where
\beqa
Q_v = -\frac{1}{8\pi V}
\Big( \frac{\kappa_{11}}{4\pi} \Big)^{2/3}
\int_{\mX} J \wedge (\text{tr} F\wedge
F-\frac{1}{2}\text{tr} R\wedge R) \; .
\label{Charge}
\eeqa
$V$ denotes the unwarped Calabi-Yau volume, $J$ the K\"ahler-form of $\mX$ and $F$ resp.~$R$ the Yang-Mills resp.~curvature two-forms on the visible boundary. Later, we will work in the smooth upstairs picture, i.e.~on the covering space $\mathbf{S}^1$ of the orbifold. The extension of the solution to the covering space is easy, noting that all three components of the metric $\hat g_{MN}$ are $\Z2$ even
\beqa
\hat g_{\mu\nu}(-x^{11}) = \hat g_{\mu\nu}(x^{11}) \; , \qquad
\hat g_{lm}(-x^{11}) = \hat g_{lm}(x^{11}) \; , \qquad \hat
g_{11,11}(-x^{11}) = \hat g_{11,11}(x^{11}) \; .
\eeqa

A characteristic feature of this warped flux compactification geometry is that the Calabi-Yau gets conformally deformed through the warp-factor and its volume decreases along $\Orb$. The volume of the Calabi-Yau on the hidden boundary becomes zero at the critical length
\beqa
L_c = 1/Q_v \; ,
\eeqa
in terms of which we can write the warp-factor on the hidden boundary as
\beqa
e^{3f(L)} = |(L_c-L)/L_c| \; ,
\eeqa
which plays an important role in what follows.

The unwarped Calabi-Yau threefold has an $SU(3)$ structure
$\{J,\Omega\}$, where $\Omega$ denotes the covariantly constant holomorphic three-form on $\mathbf{X}$. This structure is fibered along $\Orb$ and induces an $SU(3)$ structure $\{\hat v, \hat J, \hat \Omega\}$ on the 7-manifold $\mX \times \Orb$, where the real vector is given by $\hat v = e^{k(x^{11})}dx^{11}$. The following compatibility conditions must hold
\begin{alignat}{3}
i\hat \Omega \wedge \bar{\hat\Omega} &= \frac{4}{3} \hat J
\wedge \hat J \wedge \hat J \; , \\
\hat \Omega \wedge \hat J &= 0
\end{alignat}
Furthermore, $\hat v \,\lrcorner\, \hat J = \hat v \,\lrcorner\, \hat \Omega = 0$. Since $\frac{1}{3!}\hat J \wedge \hat J \wedge \hat J$ is the volume form $e^{6f(x^{11})} \sqrt{g(\mathbf{X})} dy^1 \wedge \hdots \wedge dy^6$ on the conformally deformed Calabi-Yau, it is easy to fix the $x^{11}$ dependence of $\hat J$ and $\hat \Omega$ to be
\begin{alignat}{3}
\hat J &= e^{2f(x^{11})} J \\
\hat \Omega &= e^{3f(x^{11})} \Omega \; .
\label{OmegaWarp}
\end{alignat}

\section{Vacuum Energy and Supersymmetry Breaking}

To see how the warping of the M-theory background helps in reducing the vacuum energy after supersymmetry breaking, we start by considering the perfect square potential in heterotic M-theory \cite{Horava:1996vs}
\beqa
S = -\frac{1}{2\kappa^2}
\int_{\phantom{}_{\mathbf{R^{1,3}}\times\mathbf{X}\times\mathbf{S}^1}}
\hspace{-11mm}
\Big(G-\frac{\sqrt{2}}{32\pi} \left(\frac{\kappa}{4\pi}\right)^{2/3}
\text{tr} \bar\chi \hat\Gamma^{(3)} \chi \wedge \delta_L \Big)^2
\; .
\label{HetMSquare}
\eeqa
The hat on $\hat\Gamma^{(3)}$ indicates that it contains the vielbeine of the warped metric. We are working in the upstairs picture, i.e.~on the smooth seven-manifold $\mathbf{M}^7 = \mX\times\mathbf{S}^1$ by imposing a fixed $\Z2$ symmetry on the fields along the eleventh dimension. In this picture, the Dirac one-form
\beqa
\delta_L = \left( \delta(x^{11}-L) + \delta(x^{11}+L) \right) dx^{11} \; ,
\eeqa
which is localized on the hidden boundary, includes both the hidden boundary and its $\Z2$ mirror. Minimizing the potential fixes the flux $G$ and localizes it on the hidden boundary
\beqa
G = \frac{\sqrt{2}}{32\pi} \left(\frac{\kappa}{4\pi}\right)^{2/3}
\langle \text{tr} \bar\chi \hat\Gamma^{(3)} \chi \rangle \wedge \delta_L \; .
\label{GGC}
\eeqa
$G$ is $\mathbf{Z}_2$ odd, and the two delta-functions in $\delta_L$ give it a well-defined $\mathbf{Z}_2$ symmetry. The 10-dimensional gaugino $\chi = \chi_6^\ast\otimes\chi_4 + \chi_6\otimes\chi_4^\ast$ splits into a 6- and a 4-dimensional positive chirality Weyl spinor $\chi_6$ and $\chi_4$. With $\chi_6$ being a zero mode of the Dirac operator, the condensate is only allowed to assume a vacuum expectation value on $\mathbf{X}$ of the form \cite{Dine:1985rz}
\beqa
\langle \text{tr} \bar\chi \hat\Gamma^{(3)} \chi \rangle \propto \Lambda^3 \bar{\hat\Omega}(L) + c.c.
\label{GCProp}
\eeqa
As in the case of the heterotic superstring, the fixation of $G$ in terms of $\Omega$ and its complex conjugate fixes the complex structure moduli of the compactification. Moreover, it implies that $G$ must be of Hodge type $G^{(3,0,1)}$ resp.~$G^{(0,3,1)}$. Because a $G$-flux with this Hodge type breaks supersymmetry, the dynamical minimization of the potential fixes complex structure moduli and breaks supersymmetry at zero vacuum energy.

What remains to be shown is that no conflict with the quantization of $G$ arises thanks to the warping, which we will explain next. Otherwise flux quantization would prohibit the above balancing of flux and condensate much as in the heterotic superstring case.
To this end, we note that $\hat\Gamma^{(3)} = e^{3f(L)}\Gamma^{(3)}$, when the dependence on the warp-factor is pulled out. The same warp-factor dependence for the condensate follows also from the warp-factor dependence of $\hat\Omega$, together with (\ref{GCProp}). A further warp-factor dependence of the condensate arises via the gauginos. Namely, after the reduction and truncation procedure, the 4-dimensional gaugino $\lambda$ appears with a non-canonically normalized kinetic term \cite{Lalak:1997zu}, \cite{Nilles:2004zg}
\beqa
S_{YM,h} = -\int d^4 x\sqrt{-g_4}
\Big( \frac{1}{4g_h^2}\text{tr}F_{\mu\nu}^2
+ \frac{i}{2g_h^2}\text{tr}\bar\lambda\Gamma^\mu D_\mu\lambda
\Big)
\eeqa
such that $\langle \text{tr} \bar\chi \hat\Gamma^{(3)} \chi \rangle \propto g_h^2$ \cite{Nilles:2004zg}. The 4-dimensional gaugino condensate is given by $\Lambda^3 = \langle \text{Tr} \lambda \lambda \rangle$. The second warp-factor dependence of the condensate then follows from the hidden gauge coupling's dependence on the warp-factor \cite{Curio:2000dw}, \cite{Curio:2003ur}
\beqa
g_h^2 = g_0^2 e^{-6f(L)} \; .
\eeqa
Taken together, we thus arrive at
\beqa
\langle \text{tr} \bar\chi \hat\Gamma^{(3)} \chi \rangle
= g_h^2 \big( \Lambda^3
\bar{\hat\Omega}(L) + \bar\Lambda^3{\hat\Omega}(L) \big)
= e^{-3f(L)} g_0^2 \big( \Lambda^3
\bar\Omega + \bar\Lambda^3\Omega \big) \; .
\label{Condensate}
\eeqa
The $G$-flux quantization condition in M-theory reads \cite{Witten:1996md}
\beqa
\frac{1}{\sqrt{2}} \left(\frac{4\pi}{\kappa}\right)^{2/3}
\int_{\Sigma_4} G
+ \frac{\pi}{4} \int_{\Sigma_4} p_1(\mX)
= 2\pi N \; , \qquad N \in \mathbf{Z}
\eeqa
for an arbitrary 4-cycle $\Sigma_4 \in H_4(\mX,\mathbf{Z})$ which does not connect both boundaries. $p_1(\mX)$ is the first Pontryagin class of $\mX$. In our case, $G$ has a leg in the eleventh dimension and is proportional to $\Omega$ and its complex conjugate. To obtain a non-trivial statement upon $\Sigma_4$ integration, we therefore have to choose the 4-cycle $\Sigma_4 = \Sigma_3 \times \mathbf{S}^1$, where $\Sigma_3$ is, up to a factor, the Poincar\'e dual to $\Omega$ or $\bar\Omega$. Because this 4-cycle connects both boundaries, a further boundary contribution needs in principle to be taken into account. Both, the boundary contribution and the $p_1(\mX)$ term do, however, not contribute for this cycle, as has been shown in \cite{Lukas:1997rb}. Hence we can omit the $p_1(\mX)$ term in the $G$-flux quantization condition. Applying it to our flux fixation at the potential's minimum (\ref{GGC}), we get
\beqa
e^{3f(L)} N
= \frac{\alpha_0}{8\pi}
\big( \Lambda^3\bar\Pi + \bar\Lambda^3\Pi \big) \; ,
\label{NGCBalance}
\eeqa
with gauge coupling $\alpha_0=g^2_0/(4\pi)$ and period $\Pi = \int_{\Sigma_3} \Omega$.

The new important feature, as compared to the analysis for the heterotic superstring, is the appearance of the warp-factor on the lhs which can vary between $0\le e^{3f(L)} \le 1$. This additional continuous quantity allows a balancing of the quantized $G$-flux and gaugino condensate by dynamical adjustment of the orbifold length $L$. With $\Lambda = 10^{13}\,$GeV and the period somewhat larger than the fundamental M-theory scale, $\Pi \ge \kappa^{2/3}$, we find with $\kappa^{2/9} \approx (2M_{GUT})^{-1}$ that $L$ becomes {\em stabilized very close to $L_c$}. From the relation $L\sim g_s^{2/3}$ between $L$ and string coupling $g_s$, it is clear that this adjustment is not accessible within the heterotic superstring description valid at $g_s\rightarrow 0$.

It is intriguing to have this generic mechanism for supersymmetry breaking with zero vacuum energy together with the stabilization of $L$ close to the critical length. After all, it is this point in moduli space which exhibits a realistic phenomenology, as first demonstrated in \cite{Witten:1996mz} and followed by many other works. Moreover, it is satisfying to see that moduli stabilization carried out within the 4-dimensional effective action, including the relevant non-perturbative effects, agrees with this result \cite{Curio:2001qi}, \cite{Becker:2004gw}, \cite{Curio:2006dc}. We would like to stress that this warp-factor suppression of the integer-valued flux is essentially model-independent. It is clear from the balancing equation (\ref{NGCBalance}) that the presence of the condensate acts by lowering the value of the stabilized $L$. In the absence of a condensate one has, $L=L_c$, while in its presence $L<L_c$. This feature agrees well with the effective description of the gaugino condensate via its superpotential $W\sim e^{-S/C_H+\beta T/C_H}$. The superpotential grows with $T$, the orbifold-length modulus, and thus tends to decrease the orbifold-length on energetic grounds.
Warp-factor mechanisms to achieve a critical vacuum energy in type IIB models with branes have been proposed in the past \cite{Krause:2000gp}, \cite{Krause:2000uj}. This mechanism is, however, different as the warp-factor acts exclusively on the flux part.

\section{Effective Superpotential}

We will now turn to an investigation of the effective superpotential and exhibit the same warp-factor suppression of the flux superpotential while the gaugino condensate superpotential receives no warp-factor suppression. Importantly, this allows a balancing of flux and condensate contributions in moduli stabilization. As a byproduct, we find a new derivation of the gaugino condensate's superpotential in heterotic M-theory which is free of ambiguous squares of the Dirac $\delta$-function.

The $G$-flux superpotential for the M-theory on $\Orb$ background (\ref{Back}) has been derived recently in \cite{Anguelova:2006qf} (see also \cite{Behrndt:2005im}). It reads\footnote{Our convention for $G$ differs from the $G^{AZ}$ of \cite{Anguelova:2006qf} by $G^{AZ} = e^{k(x^{11})} G$. This ensures that we are working with the same $G$ appearing in the perfect square potential which has a NS three-form decomposition $G = H\wedge dx^{11}$ rather than $G^{AZ}=H\wedge\hat v$.}
\beqa
W = \frac{1}{\kappa^2}\int_{\mathbf{X}\times \mathbf{S}^1}
\hspace{-4mm} e^{b(x^{11})+k(x^{11})} (G \wedge \hat\Omega) \; ,
\eeqa
where, again, we work in the upstairs picture. Following Horava \cite{Horava:1996vs}, we can incorporate the gaugino condensate by extending $G$ to
\beqa
G = H\wedge L \delta_L - \frac{\sqrt{2}}{32\pi} \left(\frac{\kappa}{4\pi}\right)^{2/3}
\langle \text{tr} \bar\chi \hat\Gamma^{(3)} \chi \rangle \wedge \delta_L \; ,
\label{GGCExt}
\eeqa
the first piece being the standard decomposition into the NS three-form flux $H$, which has to be located on the hidden boundary to balance the gaugino condensate. The factor $L$ converts the Delta-functions into proper dimensionless Delta-functions. We will see that this extension of $G$ readily delivers both the $H$-flux and the gaugino condensate superpotential. Since the Dirac one-form enters $G$ linearly, as opposed to quadratically when performing a reduction of the action, this derivation of the condensate's superpotential is free of the notorious $\delta_L^2$ problems.

Remembering that all three warp-factors are $\mathbf{Z}_2$ even and exploiting the supersymmetry relations (\ref{Rel1}), (\ref{Rel2}) among them, the integration over $\mathbf{S}^1$ yields
\beqa
W = e^{3f(L)} \frac{2L}{\kappa^2} \int_{\mX} H
\wedge \Omega - \frac{\alpha_0}{2\sqrt{2}\kappa^2}
\left(\frac{\kappa}{4\pi}\right)^{2/3} \Lambda^3
\int_{\mX} \bar\Omega \wedge \Omega \; ,
\eeqa
where we have used (\ref{Condensate}). Whereas the flux term keeps a background warp-factor dependence, it is canceled in the condensate term thanks to extra warp-factor dependence of the condensate. With the relation $\kappa^2
= 2L\kappa_{10}^2$ between the 11- and 10-dimensional
gravitational coupling constants, the equality, $i\int_{\mX} \Omega \wedge \bar\Omega = 8V$, and the relation, $g_0^2 V(\mX) = g_{10}^2 = 2\pi (4\pi\kappa^2)^{2/3}$, between the 4- and 10-dimensional gauge coupling \cite{Horava:1996ma}, we finally arrive at
\beqa
W = e^{3f(L)} W_H - i\sqrt{2}\Lambda^3 \; ,
\eeqa
with
\beqa
W_H = \frac{1}{\kappa_{10}^2}\int_{\mathbf{X}} H \wedge \Omega \;,
\eeqa
being the standard $H$-flux superpotential \cite{Gukov:1999ya}, \cite{Gukov:1999gr}. The condensate can be expressed as \cite{Shifman:1987ia}
\beqa
\Lambda^3 = \langle \text{Tr} \lambda\lambda\rangle
= 16 \pi^2 M_{UV}^3 e^{-f_h/C_H} \; ,
\eeqa
in terms of the gauge kinetic function, $f_h$, in the hidden sector, an ultraviolet cutoff $M_{UV}$ of the hidden gauge theory and the hidden gauge group's dual Coxeter number $C_H$ which determines the 1-loop beta-function. The gauge kinetic function is $f_h = S - \beta T$, with a parameter $\beta$ which can be expressed geometrically as
\beqa
\beta = L/L_c
\eeqa
with help of formulas presented in \cite{Becker:2004gw}. It is
thus obvious that the $\beta T$ correction becomes sizeable only in the M-theory regime. Hence, the second term delivers the usual gaugino condensate superpotential \cite{Shifman:1987ia}, \cite{Seiberg:1994bz}, \cite{Intriligator:1994jr}
\beqa
W_{GC} = - i\sqrt{2} (4\pi)^2 M^3_{UV}e^{-f_h/C_H} \; ,
\eeqa
and we can conclude that
\beqa
W = e^{3f(L)} W_H + W_{GC} \; .
\eeqa

Hence, in agreement with the perfect square analysis of the 11-dimensional action, we also find in the effective 4-dimensional action a suppression of the flux contribution by a warp-factor $e^{3f(L)}$ relative to the contribution from the gaugino condensate. From the point of view of the effective theory where
\beqa
|W_H| \sim 1 \; , \qquad |W_{GC}| \sim e^{-f_h/C_H} \ll 1
\eeqa
are typically of very different magnitudes, the warp-factor suppression of $W_H$ thus enables a balancing of both effects, a crucial requisite for moduli stabilization.

In fact the warp-factor suppression of $W_H$ alone, can also be understood directly. For this let us choose a symplectic basis of 3-cycles $A^p,B_q \in H_3(\mX,\mathbf{Z});\; p,q = 0,\hdots,h^{2,1}(\mX)$. The $H$-flux superpotential has to be evaluated at the position of the hidden boundary with $\hat\Omega(L)$ and can be expressed as
\beqa
\hat W_H = \frac{1}{\kappa_{10}^2}\int_{\mX} H \wedge \hat \Omega
= \frac{2}{(2\pi)^5\alpha'^3}\big( n_p \hat Z^p - m^q \hat F_q \big)
\eeqa
in terms of the periods $\hat Z^p = \int_{A^p} \hat\Omega$, $\hat F_q = \int_{B_q} \hat \Omega$ and the flux integers
\beqa
\frac{1}{(2\pi)^2\alpha'}\int_{A^p} H = m^p \; , \qquad
\frac{1}{(2\pi)^2\alpha'}\int_{B_q} H = n_q \; .
\eeqa
Since $\hat\Omega(L)$ scales like $e^{3f(L)}$, the periods
scale in the same way, $\hat Z^p = e^{3f(L)} Z^p$ and $\hat F_q = e^{3f(L)} F_q$. Therefore, we recognize immediately, from the direct formulation in terms of $\hat\Omega$ or the period expression, that the $H$-flux superpotential must have the warp-factor dependence
\beqa
\hat W_H = e^{3f(L)} W_H \; .
\eeqa
We would like to stress that, whereas in type IIB string compactifications the smallness of $W_H$ requires string landscape statistics plus anthropic reasoning, it arises here directly as a result of the warped background.

\bigskip
\noindent {\large \bf Acknowledgements}\\[-1mm]

\noindent We want to thank Lubos Motl and Hans Peter Nilles for related helpful discussions. This work has been supported by the DFG and the Transregional Collaborative Research Centre TRR~33 ``The Dark Universe''.

\bibliographystyle{plain}

\begin{thebibliography}{99}

\bibitem{Dine:1985rz}
  M.~Dine, R.~Rohm, N.~Seiberg and E.~Witten,
  Phys.\ Lett.\ B {\bf 156}, 55 (1985).

\bibitem{Bergshoeff:1989de}
  E.~A.~Bergshoeff and M.~de Roo,
  Nucl.\ Phys.\ B {\bf 328}, 439 (1989).

\bibitem{Ferrara:1982qs}
  S.~Ferrara, L.~Girardello and H.~P.~Nilles,
  Phys.\ Lett.\ B {\bf 125}, 457 (1983).\\
  J.~P.~Derendinger, L.~E.~Ibanez and H.~P.~Nilles,
  Phys.\ Lett.\ B {\bf 155}, 65 (1985).\\
  J.~P.~Derendinger, L.~E.~Ibanez and H.~P.~Nilles,
  Nucl.\ Phys.\ B {\bf 267}, 365 (1986).

\bibitem{Rohm:1985jv}
  R.~Rohm and E.~Witten,
  Annals Phys.\  {\bf 170}, 454 (1986).

\bibitem{Derendinger:1986}
  J.~P.~Derendinger, L.~E.~Ibanez and H.~P.~Nilles,
  Nucl.\ Phys.\ B {\bf 267}, 365 (1986).

\bibitem{Gukov:2003cy}
  S.~Gukov, S.~Kachru, X.~Liu and L.~McAllister,
  Phys.\ Rev.\ D {\bf 69}, 086008 (2004)
  [arXiv:hep-th/0310159].

\bibitem{Curio:2005ew}
  G.~Curio, A.~Krause and D.~L\"ust,
  Fortsch.\ Phys.\  {\bf 54}, 225 (2006)
  [arXiv:hep-th/0502168].

\bibitem{Curio:2000dw}
  G.~Curio and A.~Krause,
  Nucl.\ Phys.\ B {\bf 602}, 172 (2001)
  [arXiv:hep-th/0012152].

\bibitem{Krause:2001qf}
  A.~Krause,
  Fortsch.\ Phys.\  {\bf 49}, 163 (2001).

\bibitem{Curio:2003ur}
  G.~Curio and A.~Krause,
  Nucl.\ Phys.\ B {\bf 693}, 195 (2004)
  [arXiv:hep-th/0308202].

\bibitem{Witten:1996mz}
  E.~Witten,
  Nucl.\ Phys.\ B {\bf 471}, 135 (1996)
  [arXiv:hep-th/9602070].

\bibitem{Horava:1996vs}
  P.~Horava,
  Phys.\ Rev.\ D {\bf 54}, 7561 (1996)
  [arXiv:hep-th/9608019].

\bibitem{Lalak:1997zu}
  Z.~Lalak and S.~Thomas,
  Nucl.\ Phys.\ B {\bf 515}, 55 (1998)
  [arXiv:hep-th/9707223].

\bibitem{Nilles:2004zg}
  H.~P.~Nilles,
  arXiv:hep-th/0402022.

\bibitem{Witten:1996md}
  E.~Witten,
  J.\ Geom.\ Phys.\  {\bf 22}, 1 (1997)
  [arXiv:hep-th/9609122].

\bibitem{Lukas:1997rb}
  A.~Lukas, B.~A.~Ovrut and D.~Waldram,
  Phys.\ Rev.\ D {\bf 57}, 7529 (1998)
  [arXiv:hep-th/9711197].

\bibitem{Curio:2001qi}
  G.~Curio and A.~Krause,
  Nucl.\ Phys.\ B {\bf 643}, 131 (2002)
  [arXiv:hep-th/0108220].

\bibitem{Becker:2004gw}
  M.~Becker, G.~Curio and A.~Krause,
  Nucl.\ Phys.\ B {\bf 693}, 223 (2004)
  [arXiv:hep-th/0403027].

\bibitem{Curio:2006dc}
  G.~Curio and A.~Krause,
  Phys.\ Rev.\  D {\bf 75}, 126003 (2007)
  [arXiv:hep-th/0606243].

\bibitem{Krause:2000gp}
  A.~Krause,
  Nucl.\ Phys.\ B {\bf 748}, 98 (2006)
  [arXiv:hep-th/0006226].

\bibitem{Krause:2000uj}
  A.~Krause,
  JHEP {\bf 0309}, 016 (2003)
  [arXiv:hep-th/0007233].

\bibitem{Anguelova:2006qf}
  L.~Anguelova and K.~Zoubos,
  Phys.\ Rev.\ D {\bf 74}, 026005 (2006)
  [arXiv:hep-th/0602039].

\bibitem{Behrndt:2005im}
  K.~Behrndt, M.~Cvetic and T.~Liu,
  Nucl.\ Phys.\ B {\bf 749}, 25 (2006)
  [arXiv:hep-th/0512032].

\bibitem{Horava:1996ma}
  P.~Horava and E.~Witten,
  Nucl.\ Phys.\ B {\bf 475}, 94 (1996)
  [arXiv:hep-th/9603142].

\bibitem{Gukov:1999ya}
  S.~Gukov, C.~Vafa and E.~Witten,
  Nucl.\ Phys.\ B {\bf 584}, 69 (2000)
  [Erratum-ibid.\ B {\bf 608}, 477 (2001)]
  [arXiv:hep-th/9906070].

\bibitem{Gukov:1999gr}
  S.~Gukov,
  Nucl.\ Phys.\ B {\bf 574}, 169 (2000)
  [arXiv:hep-th/9911011].

\bibitem{Shifman:1987ia}
  M.~A.~Shifman and A.~I.~Vainshtein,
  Nucl.\ Phys.\ B {\bf 296}, 445 (1988)
  [Sov.\ Phys.\ JETP {\bf 66}, 1100 (1987)]. \\
  M.~A.~Shifman and A.~I.~Vainshtein,
  Nucl.\ Phys.\ B {\bf 359}, 571 (1991).

\bibitem{Seiberg:1994bz}
  N.~Seiberg,
  Phys.\ Rev.\ D {\bf 49}, 6857 (1994)
  [arXiv:hep-th/9402044].

\bibitem{Intriligator:1994jr}
  K.~A.~Intriligator, R.~G.~Leigh and N.~Seiberg,
  Phys.\ Rev.\ D {\bf 50}, 1092 (1994)
  [arXiv:hep-th/9403198].

\end{thebibliography}

\end{document}